\documentclass[epsfig,floats,prl,twocolumn,preprintnumbers,floatfix,superscriptaddress]{revtex4-1}
\usepackage{graphicx}
\usepackage[latin2]{inputenc}
\usepackage{amsmath}
\usepackage{amssymb}
\usepackage{bm}
\usepackage{color}

\newcommand{\pf}{p\!_{_\infty}}
\newcommand{\lp}{\ell_p}
\newcommand{\lpf}{\ell_{p\!_{_\infty}}\!\!}
\newcommand{\lpi}{\ell_{p\!_{_\circ}}}
\newcommand{\lpa}{\langle\ell_p\rangle}
\newcommand{\kpv}{\kappa_{_{pv}}}
\newcommand{\kvv}{\kappa_{_{vv}}}

\begin{document}
\title{Persistence-Speed Coupling Enhances the Search Efficiency of Migrating Immune Cells}
\author{M. Reza Shaebani}
\email{corresponding author. shaebani@lusi.uni-sb.de}
\affiliation{Department of Theoretical Physics, Saarland University, 66123 
Saarbr\"ucken, Germany}
\affiliation{Center for Biophysics, Saarland University, 66123 Saarbr\"ucken, Germany}
\author{Robin Jose}
\affiliation{Department of Theoretical Physics, Saarland University, 66123 
Saarbr\"ucken, Germany}
\author{Ludger Santen}
\affiliation{Department of Theoretical Physics, Saarland University, 66123 
Saarbr\"ucken, Germany}
\affiliation{Center for Biophysics, Saarland University, 66123 Saarbr\"ucken, Germany}
\author{Luiza Stankevicins}
\affiliation{INM-Leibniz Institute for New Materials, 66123 Saarbr\"ucken, Germany}
\author{Franziska Lautenschl\"ager}
\affiliation{Center for Biophysics, Saarland University, 66123 Saarbr\"ucken, Germany}
\affiliation{INM-Leibniz Institute for New Materials, 66123 Saarbr\"ucken, Germany}
\affiliation{Department of Experimental Physics, Saarland University, 66123 
Saarbr\"ucken, Germany}

\begin{abstract}
Migration of immune cells within the human body allows them to fulfill 
their main function of detecting pathogens. Adopting an optimal navigation 
and search strategy by these cells is of crucial importance to achieve 
an efficient immune response. Analyzing the dynamics of dendritic cells 
in our in vitro experiments reveals that the directional persistence of 
these cells is highly correlated with their migration speed, and that the 
persistence-speed coupling enables the migrating cells to reduce their 
search time. We introduce theoretically a new class of random search 
optimization problems by minimizing the mean first-passage time (MFPT) 
with respect to the strength of the coupling between influential 
parameters such as speed and persistence length. We derive an analytical 
expression for the MFPT in a confined geometry and verify that the 
correlated motion improves the search efficiency if the mean persistence 
length $\lpa$ is sufficiently shorter than the confinement 
size. In contrast, a positive persistence-speed correlation even increases 
the MFPT at long $\lpa$ regime, thus, such a strategy 
is disadvantageous for highly persistent active agents.  
\end{abstract}

\maketitle
A successful immune response crucially depends on its first steps: finding 
harmful pathogens. In general, search and transport efficiency of random 
processes have been quantified by observables such as the diffusivity of 
randomly moving particles \cite{Bertrand18}, the reactivity of transport-limited 
chemical reactions \cite{Loverdo08}, the cover time to visit all sites of 
a confined domain \cite{Chupeau14,Chupeau15}, or often by the mean first-passage 
time (MFPT) that a searcher needs to find a target \cite{Benichou11,Redner01}. 
Optimal search strategies considered so far minimize the MFPT or equivalently 
the cover time with respect to one of the key parameters of the problem. This 
can be either a structural property of the environment in which the particle 
moves \cite{Schwarz16,Shaebani18} or a parameter of the stochastic motion 
(e.g., the persistency in active random searches \cite{Tejedor12}, the 
resetting rate in diffusion processes with stochastic resetting to the 
initial position \cite{Evans11,Kusmierz14}, the ratio between the durations 
of diffusive and directed motion in intermittent searches \cite{Benichou11,
Benichou05,Loverdo09}, or the speed of the searcher when passing over a 
target location \cite{Campos12}). However, the influential factors governing 
the search efficiency are correlated in general. For instance, a universal 
coupling between migration speed and directional persistence has been 
recently reported for various cell lines mediated by retrograde actin 
flows \cite{Maiuri15}. Alternative optimal search strategies for such 
correlated stochastic processes need to be developed.

\begin{figure}[b]
\centering
\includegraphics[width=0.48\textwidth]{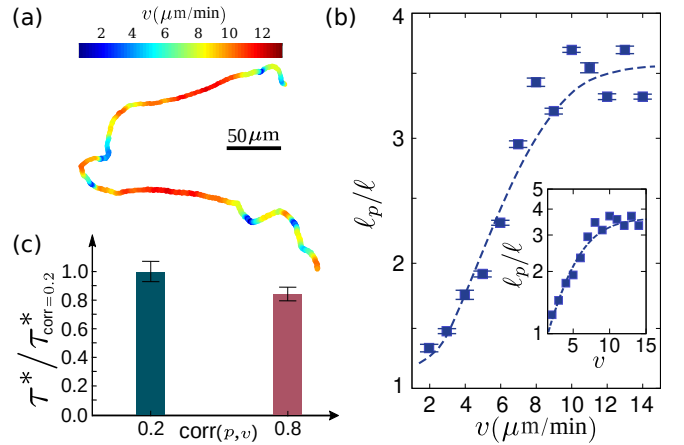}
\caption{(a) Sample cell trajectory, color coded with respect to speed. 
(b) Cell persistence length $\ell_p$ (scaled by the mean distance $\ell$ 
between successive recorded positions) in terms of migration speed $v$. 
The dashed line shows the fit via Eq.\,(\ref{Eq:lp}). Inset: Log-lin 
plot of $\ell_p{/}\ell$ vs $v$.  (c) Comparison between the conditional 
MFPT $\tau^*$ of two categories of cells with low and high $p{-}v$ 
correlation.}
\label{Fig:1}
\end{figure}

Adopting an efficient search and navigation strategy is of particular 
importance in biological systems as, for example, in search for specific 
target sites over a DNA strand by proteins \cite{Lomholt05,Elf07,Bauer12}, 
escape through small absorbing boundaries and targeted intracellular 
transport \cite{Schwarz16,Schuss07}, delivery of chemical signals in 
neurons \cite{Bressloff07,Jose18,Fedotov08}, bacterial swimming and 
chemotaxis \cite{Najafi18,Benichou11,Wadhams04,PerezIpina19}, and 
animal foraging \cite{Campos12,Bartumeus08,Oshanin09}. It is often 
hypothesized that the motility of mammalian cells enables them to 
effectively fulfill their biological functions. Migration of immune 
cells \cite{Harris12,Lavi16,Chabaud15}, which is expected to be 
optimized in the course of evolution to achieve an efficient immune 
response, is of particular interest. Nevertheless, the optimality 
of the search for pathogens and other targets by immune cells has 
neither been precisely verified nor systematically studied. 
Understanding the mechanisms of adaptive search and clearance in 
the immune system opens the way toward more effective cancer 
immunotherapies and vaccine design. 

Here we consider theoretically a correlated stochastic 
process and introduce, for the first time, a new class of optimal 
search strategies based on tuning the strength of coupling between 
key parameters. Inspired by the observed correlations in the dynamics 
of dendritic cells \cite{Wu14,Maiuri15}, we consider the correlation 
between the migration speed $v$ and directional persistence $p$ of 
the searcher. The optimization is achieved by analytically calculating 
the MFPT and minimizing it with respect to the strength $\kpv$ of 
$p{-}v$ coupling. The success of the scheme in improving the MFPT 
nontrivially depends on the ratio between the mean persistence length 
$\lpa$ of the searcher and the system size $L$; in the regime $\lpa
{\ll}L$ ($\lpa{\sim}L$), the correlated motion is advantageous 
(disadvantageous) for reducing the search time. We experimentally 
investigate the dynamics of dendritic cells (responsible for tissue 
patrolling and antigen capture \cite{Heuze13,Chabaud15}) and 
expectedly observe a significant persistence-speed correlation 
\cite{Wu14,Maiuri15} (see Fig.\,\ref{Fig:1}). Our data analysis 
also reveals an interesting inverse dependence of the MFPT on the 
coupling strength, in agreement with our analytical predictions 
for the low persistence regime. 

{\it Migration of dendritic cells.---} To study the dynamics of migrating 
cells in our in vitro experiments, we tracked the 2D motion of Murine bone 
marrow-derived immature dendritic cells with typical size of nearly ten 
micrometers. The motion was confined between the cell culture dish and a 
roof held by microfabricated pillars made out of Polydimethylsiloxane (PDMS) 
as described in \cite{Leberre14} at a height of 3$\,\mu$m. Both surfaces 
were coated with PLL-PEG (0.5 mg/mL), a non-adhesive material to exclude 
movement by cell adhesion. The cell concentration was low enough to treat 
the cells as non-interacting. Cell nuclei were stained with Hoechst 34580 
(200 ng/mL for 30 min) (Sigma Aldrich, St Louis, USA) and migration was 
recorded by epifluorescence microscopy for at least 6h at 37$^\circ$ 
with a camera of 6.5$\,\mu$m pixel size and sampling rate of 20 frames/h. 

A typical cell trajectory is shown in Fig.\,\ref{Fig:1}(a), evidencing 
that the path is more straight when the migration speed is higher. We 
quantify the cell persistence--- the ability of the cell to maintain its 
current direction of motion--- by $p\,{=}\cos\theta$ with $\theta$ being 
the orientational change at each recorded position \cite{Shaebani14,
Sadjadi15,Burov13}, from which the instantaneous persistence length $\ell_p$ 
can be estimated as $\lp{=}\frac{-\ell}{\ln|p|}$ ($\ell$ is the mean 
distance between two successive recorded positions) \cite{Shaebani19}. 
The leading contribution in the limit of small $\theta$ goes as $\mathcal{O}
(\frac{\ell}{1{-}p})$. After averaging over all trajectories and speed 
binning intervals of $\Delta v{=}1\,\mu\text{m}{/}\text{min}$, we observe 
a clear coupling between the cell persistence $p$ and the migration speed 
$v$, which can be fitted by an exponential saturation $p\,=\pf(1{-}
\text{e}^{-\gamma v})$, with $\pf{\approx}\,0.7$ and $\gamma\,{\approx}\,
0.3$. The behavior of $\lp$ is well fitted by a logistic function
\begin{eqnarray}
\lp = \frac{\lpf}{1+(\displaystyle\frac{\lpf\!-\lpi}{\lpi})
\text{e}^{-\gamma v}},
\label{Eq:lp}
\end{eqnarray}
where $\lpi$ is the persistence length of a nonpersistent motion and 
$\lpf{\simeq}\,\frac{\lpi}{1{-}\pf}$ [Fig.\,\ref{Fig:1}(b)]. $\lp$ 
initially grows exponentially as $\lp\,{\propto}\,\text{e}^{\gamma v}$ 
\cite{Maiuri15} but eventually saturates to $\lpf$ at high speeds. 
To describe the overall coupling strength for individual cells we 
calculate the $p{-}v$ correlation coefficient $\text{corr}(p,v){=}
\frac{\text{cov}(p,v)}{\sigma_p\sigma_v}$ for each cell. When averaged 
over all trajectories, a strong correlation around $0.9$ is obtained. 

\begin{figure}[b]
\centering
\includegraphics[width=0.48\textwidth]{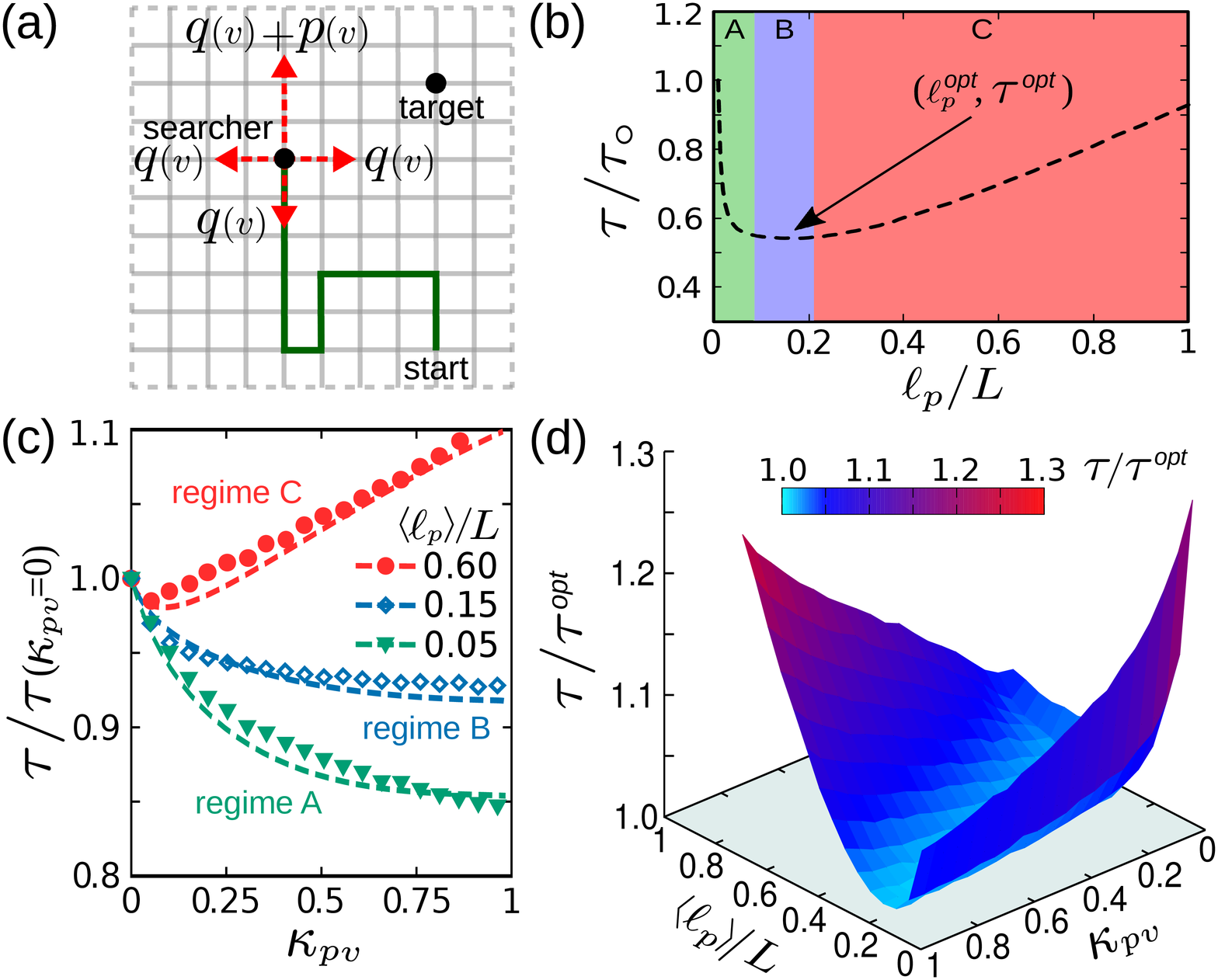}
\caption{(a) Sketch of the correlated persistent random search on a 
square lattice. (b) MFPT of a constant $\lp$ and $v$ process, scaled 
by the MFPT of a diffusive searcher, vs $\lp$ normalized by $L$ ($L{=}
200$). (c) MFPT of a correlated $p{-}v$ process, scaled by the MFPT 
when $\kpv{=}0$, vs the coupling strength $\kpv$. Each curve belongs 
to a different regime of $\lpa$ shown in panel (b). The dashed lines 
represent analytical predictions and the symbols are simulation 
results. (d) MFPT scaled by the optimal search time of the constant 
$\lp$ process vs $\kpv$ and scaled $\lpa$.}
\label{Fig:2}
\end{figure}

The key question is whether such a correlated random motion helps the 
immune cells to improve their search efficiency. To answer this, we 
selected two subpopulations of cells with distinct mean correlation 
coefficients $0.2\pm0.05$ and $0.8\pm0.05$. By calculating the 
conditional MFPT $\tau^*$--- i.e.\ over successful trials to reach 
a random hidden target--- per unit area for each category (scaled 
by their mean speeds) and various target sizes we obtain $10{-}15\%$ 
lower search times at higher correlations, as shown in Fig.\,\ref{Fig:1}(c). 
In order to understand these MFPT results we develop a stochastic 
model for correlated persistent search in the following, and prove 
that $p{-}v$ coupling strategy is only beneficial for relatively 
weak persistencies, as in the case of dendritic cells. 

{\it Correlated persistent search model.---} We consider a discrete-time 
persistent random walk on a two-dimensional square lattice of size $L$ with 
periodic boundary conditions [Fig.\,\ref{Fig:2}(a)]. At each time step, 
the searcher moves $v$ steps drawn from a speed distribution $f(v)$. It 
either continues along the previous direction of motion with probability 
$q{+}p$ or chooses a new direction, each with a probability $q$, so that 
$4q{+}p{=}1$. The persistency parameter $p$ (and thus $q$) depends on 
the instantaneous speed $v$ and ranges from 0 (ordinary diffusion) to 1 
(ballistic motion). The instantaneous persistence length can be obtained as $\lp
{=}\sum\limits_{\ell{=}1}^{\infty}\ell(q{+}p)^{\ell{-}1}(1{-}q{-}p){=}
\frac43\frac{1}{1{-}p(v)}$. 

Assuming that a single target of one lattice-unit size is located at 
${\bm r\!}_{_T}$ (equivalent to regularly spaced targets on an 
infinite plane with $\frac{1}{L^2}$ density), we introduce $\tau({\bm 
r},v,\sigma)$ as the MFPT of reaching the target starting at position 
${\bm r}\,({\neq}{\bm r\!}_{_T})$ with speed $v$ and orientation $\sigma
\!\in\!\{\rightarrow,\leftarrow,\uparrow,\downarrow\}$. The evolution 
of $\tau({\bm r},v,\sigma)$ can be described by the following backward 
master equation
\begin{equation}
\begin{split}
&\tau({\bm r},v,\rightarrow){=}\!\int \!\!\text{d}v' f(v') \Big[(q{+}p)\;
\tau({\bm r}{+}v\hat{\bf{i}},v',\rightarrow){+}\\ 
&q\;\tau({\bm r}{-}v\hat{\bf{i}},v',\leftarrow){+}q\;\tau({\bm r}{+}v
\hat{\bf{j}},v',\uparrow){+}q\;\tau({\bm r}{-}v\hat{\bf{j}},v',\downarrow)
{+}1\Big],
\end{split}
\label{Eq:BMEr}
\end{equation}
and similar master equations for $\tau({\bm r},v,\leftarrow)$, $\tau({\bm r},
v,\uparrow)$ and $\tau({\bm r},v,\downarrow)$. The possible velocities are 
limited to the integer values $v'{\in}[0,v_\text{max}]$, which are supposed to be 
equally probable for simplicity. By introducing the Fourier transform 
$\tau({\bm k},v,\sigma){=}\sum\limits_{\bm r}\tau({\bm r},v,\sigma)\,
\text{e}^{-\text{i}{\bm k}\cdot{\bm r}}$ and using $\int\!\text{d}v f(v) 
\,\tau({\bm k},v,\sigma)=\tau({\bm k},\sigma)$ for a uniform distribution 
$f(v)$, after some calculations we obtain
\begin{equation}
\tau({\bm k},\sigma){=}\frac{F(k){+}S\Big(\delta({\bm k}){-}\text{e}^{-
\text{i}{\bm k}\cdot{\bm r\!}_{_T}}\Big)}{1{-}B_\sigma(k)},
\label{Eq:BMEk}
\end{equation}
with $B_\sigma(k){=}\frac1L\!\!\!\sum\limits_{v{=}0}^{v_\text{max}}\!\!p(v)
\text{e}^{\text{i}vk_\sigma}\!$, $F(k){=}\frac1L\!\!\sum\limits_v 
p(v)\!\!\sum\limits_\sigma\!\text{e}^{\text{i}vk_\sigma}\tau({\bm k},
\sigma)$, $S{=}L^2$, and $k_\sigma{\in}\{{\bm k}\cdot\hat{\bf{i}},
-{\bm k}\cdot\hat{\bf{i}},{\bm k}\cdot\hat{\bf{j}},-{\bm k}\cdot
\hat{\bf{j}}\}$. Next we multiply Eq.\,(\ref{Eq:BMEk}) by 
$\text{e}^{\text{i}vk_\sigma}$ and sum over $\sigma$ and $v$ to 
derive a closed expression 
\begin{equation}
F(k){=}\frac{A(k)\,S\,\Big(\delta({\bm k}){-}\text{e}^{-\text{i}{\bm k}
\cdot{\bm r\!}_{_T}}\Big)}{1{-}A(k)}.
\label{Eq:Fk}
\end{equation}
Here $A(k){=}\frac1L\!\sum\limits_v p(v)\!\sum\limits_\sigma\!
\frac{\text{e}^{\text{i}vk_\sigma}}{1{-}B_\sigma(k)}$. Inserting 
$F(k)$ into Eq.\,(\ref{Eq:BMEk}) and averaging over all directions 
$\sigma$ then yields
\begin{equation}
\tau({\bm k}){=}\frac{C(k)\,S\,\Big(\delta({\bm k}){-}\text{e}^{-
\text{i}{\bm k}\cdot{\bm r\!}_{_T}}\Big)}{1{-}A(k)},
\label{Eq:Tk}
\end{equation}
where $C(k){=}\frac14\!\sum\limits_\sigma\!\frac{1}{1{-}B_\sigma(k)}$. 
Finally, we apply the inverse Fourier transform (with the components 
of available modes being $k_i{=}\frac{2\pi n_i}{L}$, $n_i{\in}[0,L{-}1]$) 
and numerically average over all possible starting positions $\bm r$ 
to obtain the overall MFPT $\tau$. 

In the case of constant persistence and speed, the results of a 
single-state persistent random search \cite{Tejedor12} are recovered, 
where the MFPT shows a minimum $\tau^{\text{opt}}$ at an optimal 
persistence length $\lp^{\text{opt}}$; see Fig.\,\ref{Fig:2}(b). 
The optimal value $\frac{\lp^{\text{opt}}}{L}$ slightly decreases 
with increasing $L$. For correlated random searches, we consider a 
linear relation between $\lp$ and $v$ for simplicity, corresponding 
to an expansion of Eq.\,(\ref{Eq:lp}) up to the first order term in 
$v$. We use
\begin{equation}
\displaystyle\frac{\lp}{\lpa}{=}\kpv(\tilde{v}{-}1){+}1,
\label{Eq:lpapp}
\end{equation}
with $\tilde{v}$ being the scaled speed $\tilde{v}{=}\frac{v}{\langle 
v \rangle}$ and $\kpv$ the strength of persistence-speed coupling. 
The persistence length $\lp$ equals $\lpa$ for zero coupling 
coefficient and ranges within $[0,2\lpa]$ for $\kpv{=}1$. By inserting 
the resulting persistence parameter $p(v)$ in the above formalism, we 
obtain $\tau(\lpa, \kpv)$. We checked that using Eq.\,(\ref{Eq:lp}) 
instead of Eq.\,(\ref{Eq:lpapp}) yields qualitatively analogous 
results to those reported in the following.

\begin{figure}[b]
\centering
\includegraphics[width=0.48\textwidth]{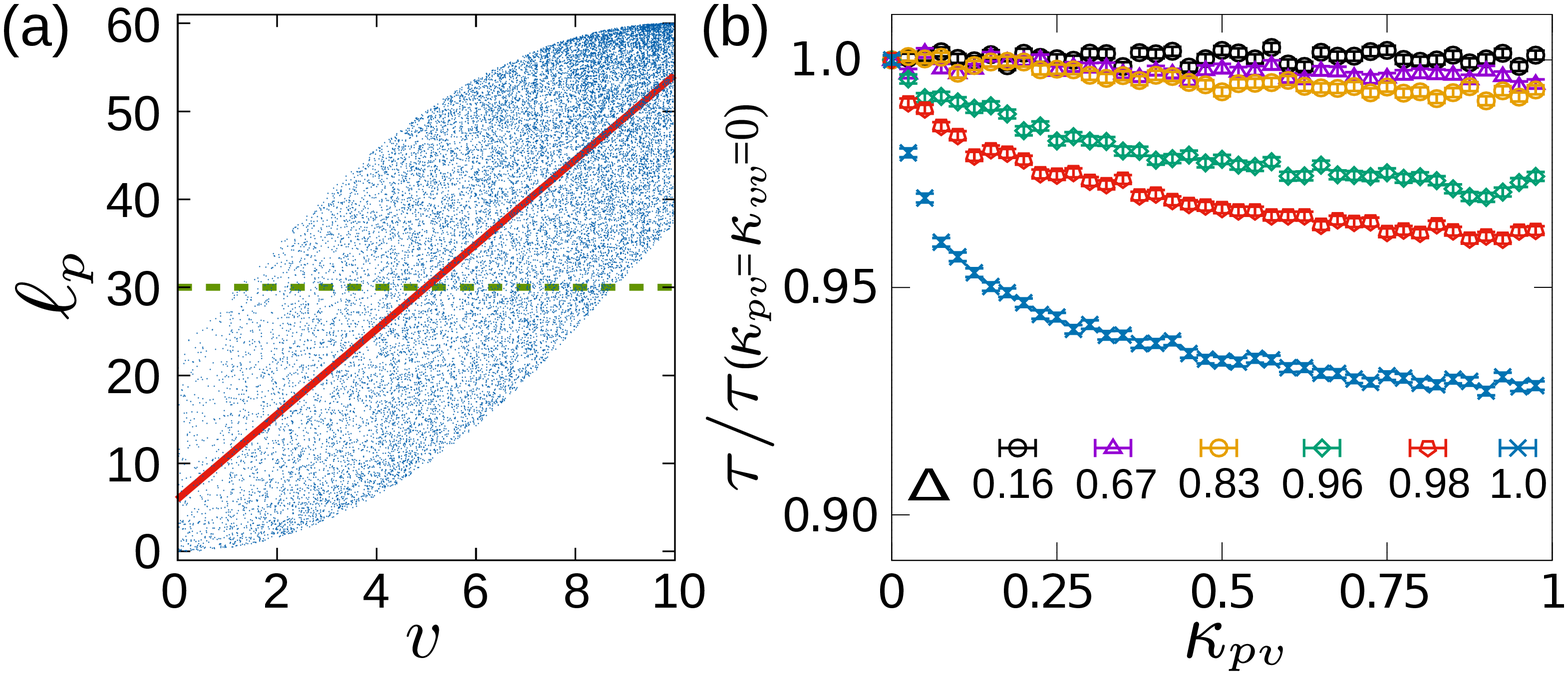}
\caption{(a) Example of correlated $(v,\lp)$ pairs generated in 
simulations with the coupling strength $\kpv{=}0.8$, drawn 
from a distribution with $\lpa{=}30$ (dashed line) and width 
$\Delta{=}1$. The solid line represents the $p{-}v$ coupling 
according to Eq.\,(\ref{Eq:lpapp}). $L{=}200$. (b) Influence 
of the distribution width $\Delta$ on the MFPT as a function 
of the coupling strength $\kpv$.}
\label{Fig:3}
\end{figure}

\begin{figure*}[t!]
\centering
\includegraphics[width=0.95\textwidth]{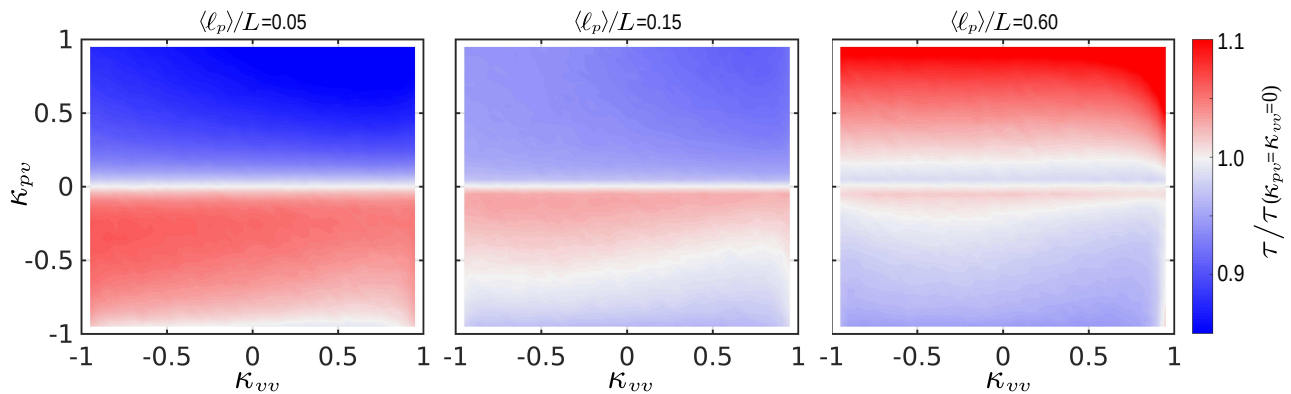}
\caption{Phase diagram of the MFPT in $(\kvv,\kpv)$ plane for different 
values of mean persistence length $\lpa$. The search time of an uncorrelated 
$p{-}v$ process in the absence of speed autocorrelations $\tau(\kvv{=}
\kpv{=}0)$ is taken as the reference for comparison in each panel. The 
color intensity indicates the deviation from the uncorrelated-search 
time, with red (blue) reflecting a decrease (increase) in the search 
efficiency.}
\label{Fig:4}
\end{figure*}

{\it Combined effects of $\lpa\!$ and $\kpv$ on search efficiency.---} 
Interestingly, Fig.\,\ref{Fig:2}(c) reveals different dependencies 
of the MFPT on the coupling strength $\kpv$ for choices of 
$\lpa$ taken from low, intermediate, and high persistence-length 
regimes A, B, C, as specified in Fig.\,\ref{Fig:2}(b). While 
$\tau$ is a decreasing function of $\kpv$ at low $\lpa$, the search 
efficiency at high mean persistence lengths even reduces with 
increasing $\kpv$. Compared with the optimal choice of the constant 
persistence length strategy, the $p{-}v$ correlated search is always 
less efficient but approaches the search time $\tau^{\text{opt}}$ 
of the former strategy at $\lpa$ values around $\lp^{\text{opt}}$; 
see Fig.\,\ref{Fig:2}(d). Note that even at $\kpv{=}\,0$ the two 
strategies are not equivalent as the velocity is a variable 
quantity in the correlated search strategy (uniformly distributed 
within $[0,2\langle v \rangle]$). The fact that the search time 
for the optimal choice of constant persistence length $\lp^{\text{opt}}$ 
is the absolute minimum over all correlated and uncorrelated persistent 
searches provides a qualitative explanation for the observed behavior 
in correlated random searches; inducing $p{-}v$ coupling at low $\lpa$ 
regime A helps to effectively increase the persistence of motion i.e.\ 
toward $\lp^{\text{opt}}$. This is in sharp contrast to the high $\lpa$ 
regime C where the increase of effective persistence length by $p{-}v$ 
correlations drags it away from $\lp^{\text{opt}}$ leading to a less 
efficient search. In the plateau regime B the $p{-}v$ coupling is 
expectedly less influential. We can analytically verify a distinct 
dependency of $\tau$ on $\kpv$ at two extreme regimes $\lpa{\rightarrow}0$ 
and $\lpa{\rightarrow}L$: Up to the leading order term, $\tau$ increases 
linearly with $\kpv$ at high persistency as $\tau(\lpa{\rightarrow}L)
{\sim}\frac{1}{1{-}p}{=}a_1\lpa\kpv\!{+}a_2$; however, it decreases 
inversely with $\kpv$ when the persistency is extremely low, where it 
can be shown that $\tau(\lpa{\rightarrow}0){\sim}\frac{1{-}p}{1{+}
p}{=}\frac{1}{b_1\lpa\kpv\!{+}b_2}$ ($a_1,b_1{>}0$).  

{\it Speed autocorrelation.---} So far we analytically obtained the 
MFPT in the presence of persistence-speed correlation for a randomly 
varying speed at each time step. However, the successive instantaneous 
speeds can be correlated in general such that the searcher experiences 
a rather smooth speed change over time. For instance, we obtain a 
positive speed autocorrelation coefficient $\kvv{\approx}\,0.3$ for 
the dendritic cells in our experiments. To incorporate the speed 
autocorrelation in our analytical approach, one should replace the 
speed distribution $f(v')$ in the master equation~(\ref{Eq:BMEr}) 
with the probability distribution of speed change $f(v{-}v')$. 
Analytical determination of the MFPT for autocorrelated speed however 
appears to be intractable; thus, we resort to Monte Carlo simulations 
to generate the desired stochastic motion.

In our simulations, we use the \textit{sum-of-uniforms} algorithm 
\cite{Lakhan81,Willemain93,Chen05} to correlate speed and persistence 
length and to include speed autocorrelation. The algorithm allows for 
inducing a certain degree of stochasticity in the resulting $v$ and 
$\lp$ values, which is controlled by an additional parameter $\Delta$. 
At each time step, first a new speed is chosen from a distribution 
around the current speed, which generates the demanded speed 
autocorrelation $\kvv$. Then a new $\lp$ is chosen from a uniform 
distribution of $\lp$ values around the value determined by the 
$p{-}v$ coupling strength $\kpv$ and the local speed $v$ according 
to Eq.\,(\ref{Eq:lpapp}) [red line in Fig.\,\ref{Fig:3}(a)]. This 
results in the cloud of blue dots in the figure. The parameter 
$\Delta\,{\in}[0,1]$ tunes the actual slope of the cloud (the upper 
limit is however set by $\kpv$) and allows for $\pm\Delta\lpa$ 
overall fluctuations. As shown in Fig.\,\ref{Fig:3}(b), $\tau$ 
approaches the MFPT of uncorrelated motion by decreasing the 
scattering parameter $\Delta$. Here we show the simulation 
results for $\Delta{=}1$ corresponding to the widest overall 
range of persistence length $[0,2\lpa]$. Once the new $v$ and 
$\lp$ are determined, we extract the instantaneous persistence 
$p$ of the searcher and move it $v$ sub-steps within one time 
step by allowing it to change the direction of motion after 
each sub-step according to the persistence probability $p$.

The results of uncorrelated speeds $\kvv{=}\,0$ in different regimes 
of $\lpa$ are shown in Fig.\,\ref{Fig:2}(c); the agreement between 
analytical predictions and simulation results is satisfactory. When 
speed autocorrelations are switched on, we find that the trends 
reported in Fig.\,\ref{Fig:2}(c) remain qualitatively valid. $\kvv$ 
plays a relatively insignificant role in determining the search time, 
while $\lpa$ and $\kpv$ are influential factors. We extend the range 
of correlation coefficients $\kpv$ and $\kvv$ to negative values for 
anti-correlated dynamics. Figure~\ref{Fig:4} summarizes the results 
in a phase diagram of search times in ($\kvv,\kpv$) plane. $\tau$ 
shows only modest dependence on $\kvv$ (subtle color intensity changes 
along horizontal lines) but variation of $\kpv$ may cause even up 
to $25\%$ changes in the search time. Another point is that inducing 
$p{-}v$ anticorrelation reduces the effective persistence of motion, 
thus, acts in the opposite direction, i.e.\ it improves the search 
time in regime C while leads to an increased search time in regime A. 

Immature dendritic cells are located in the interstitial space of 
peripheral tissues. In the skin, for example, the dermal dendritic 
cells are present in a high density of a few hundred cells per 
$\text{mm}^2$ \cite{Ng08}. If each dendritic cell patrols, on 
average, an area of linear size $L{\sim}100\,\mu\text{m}$ with a 
persistence length of less than $10\,\mu\text{m}$ (for typical 
speeds of $3{-}4\,\mu\text{m}{/}\text{min}$ and assuming even a 
high persistence $p\,{\approx}\,0.7$ before reaching the $p{-}v$ 
saturation regime), then these cells belong to the weakly persistent 
regime A in Fig.\,\ref{Fig:2}(b) (indeed regime A is even more 
extended to right for such small patrolling areas). In small 
intrapulmonary airways, the density of dendritic cells is less 
than a hundred per $\text{mm}^2$ in the absence of inflammation 
\cite{SchonHegrad91}. In such regions, each cell is responsible 
for patrolling a larger area and the corresponding relative 
persistence length in Fig.\,\ref{Fig:2}(b) further shifts to the 
left in zone A. As a result, $p{-}v$ correlations are beneficial 
for immature dendritic cells to improve the search efficiency in 
various biological environments. 
  
In summary, our study suggests improving the search efficiency 
of an active agent by inducing persistence-speed correlations 
and/or speed autocorrelations. Our key finding is that a correlated 
random motion is not necessarily an optimal search strategy in 
general; it is advantageous for dendritic cells moving with a 
persistence length much smaller than the size of the environment, 
however, highly persistent active agents should even adopt an 
anticorrelation between their speed and directional persistence 
to reduce their search time. By optimizing the search efficiency 
with respect to the strength of coupling between influential 
parameters, we introduced a new class of random search optimization 
problems with broad application to correlated stochastic processes 
such as chemotaxis and chemokinesis dynamics.   

This work was funded by the Deutsche Forschungsgemeinschaft (DFG) 
through Collaborative Research Center SFB 1027. We would like to 
thank A.~M. Lennon Dum\'enil for support with the dendritic cell 
system and Raphael Voituriez for discussions. M. R. S. and R. J. 
contributed equally.

\bibliography{Refs-Search}

\end{document}